\documentclass[12pt]{article}
\input epsf.sty
\def\ZZZ{{\hbox{ Z\kern-1.6mm Z}}}
\newcommand{\bet}{\sigma}
\newcommand{\bbb}{{\bar b}}
\newcommand{\gam}{\tau}
\newcommand{\eps}{\epsilon}
\newcommand{\ra}{\rangle}
\newcommand{\la}{\langle}

\newcommand{\tl}{\wt\lambda}

\newcommand{\BB}{{\cal B}}

\newcommand{\GG}{{\cal G}}
\newcommand{\AAA}{{\cal A}}
\newcommand{\FF}{{\cal F}}

\newcommand{\CC}{{\cal C}}

\newcommand{\wt}{\widetilde}
\newcommand{\wh}{\widehat}

\newcommand{\TT}{{\cal T}}
\newcommand{\bg}{\bar g}

\newcommand{\bG}{\bar G}
\newcommand{\bT}{\bar \Theta}
\newcommand{\SSS}{{\cal S}}

\newcommand{\be}{\begin{equation}}
\newcommand{\ee}{\end{equation}}
\newcommand{\ben}{\begin{eqnarray}\displaystyle}
\newcommand{\een}{\end{eqnarray}}
\newcommand{\refb}[1]{(\ref{#1})}
\newcommand{\p}{\partial}
\newcommand{\sectiono}[1]{\section{#1}\setcounter{equation}{0}}

\def\one{{\hbox{ 1\kern-.8mm l}}}
\def\zero{{\hbox{ 0\kern-1.5mm 0}}}

\begin{document}
{}~
{}~
\hfill\vbox{\hbox{hep-th/0208142} \hbox{MRI-O-020801}
}\break

\vskip .6cm

\centerline{\Large \bf
Decay of Unstable D-branes with Electric Field}

\medskip

\vspace*{4.0ex}

\centerline{\large \rm
Partha Mukhopadhyay$^a$ and Ashoke Sen$^{a,b}$ }

\vspace*{4.0ex}

\centerline{\large \it ~$^a$Harish-Chandra Research
Institute}

\centerline{\large \it  Chhatnag Road, Jhusi,
Allahabad 211019, INDIA}

\medskip

\centerline{\large \it ~$^b$Department of Physics, Penn State University}

\centerline{\large \it University Park,
PA 16802, USA}

\centerline{E-mail: partha@mri.ernet.in, asen@thwgs.cern.ch,
sen@mri.ernet.in}

\vspace*{5.0ex}

\centerline{\bf Abstract} \bigskip

Using the techniques of two dimensional conformal field theory we
construct time dependent classical solutions in open string theory
describing the decay of an unstable D-brane in the presence of background
electric field, and explicitly evaluate the time dependence of the energy
momentum tensor and the fundamental string charge density associated with
this solution.  The final decay product can be interpreted as a
combination of stretched fundamental strings and tachyon matter.

\vfill \eject

\baselineskip=16pt

\tableofcontents

\sectiono{Introduction and Summary} \label{s1}

It has been noticed in various works
\cite{fradkin,abouel,burgess,0005015,0005040,0006085,0205078} that
switching on an electric field on the world-volume of a D-brane
leads to various new physical effects which are absent in the
magnetic case\cite{fradkin,9711165,9908142}. In particular, unlike
the magnetic case the strength of the electric field can not be
increased beyond a critical value keeping the open string theory
stable.
The solutions carrying electric flux\cite{9704051,9901159,0009061,0010240} 
are particularly
interesting in the tachyon vacuum of an
unstable
D-brane\cite{soliton,sft}, since in this
vacuum the full Poincare invariance of the bulk is expected to be restored
and all the perturbative degrees of freedom are
unphysical
\cite{0012251}. It was shown in \cite{0009061} that although the
effective action proposed in \cite{9909062} vanishes at the tachyon 
vacuum, the system admits a
well-defined Hamiltonian
description, and has classical solutions describing electric flux tubes
carrying fundamental string charge.
The classical dynamics of these flux tubes is described by Nambu-Goto
action, and possesses the full Poincare invariance of the
bulk\cite{0009061,0010240}.

A related development in the study of tachyon dynamics has been the
construction of
time dependent classical solutions
representing the
decay of an unstable D-brane as the tachyon rolls down towards the
minimum of the
potential\cite{0202210,0203211,0203265,0204143,0207105}\footnote{
See \cite{0207107} for
the study of time dependent solutions representing rolling of tachyons
in p-adic string theory,  ref.\cite{0208028} for a time
dependent solution in cubic string field theory,
\cite{cosmo} for related cosmological applications and
\cite{0205085} for other related works including discussions in the
context of background independent string field
theory. Earlier attempts at cosmology involving rolling tachyon were
made in \cite{earl}.}.
The solution is constructed by perturbing the boundary
conformal field theory (BCFT) describing the original D-brane by an
exactly
marginal deformation. The strength $\tl$ of the perturbation labels the
initial value of
the tachyon $T$. The
perturbed
BCFT carries all the information about this one parameter family of
rolling tachyon solutions labelled by the initial position of $T$. For
example the corresponding boundary state
gives information about the time evolution of various closed string
sources, {\it e.g.} the energy-momentum tensor. In particular, it was
shown
in \cite{0203265} that if we displace the tachyon towards the
(local) minimum of
the potential and let it roll,
the system asymptotically evolves
to a pressure-less gas with non-zero energy density confined in the
initial D-brane world-volume. It was also shown that at $\tl=1/2$, which
corresponds to placing the tachyon at the minimum of the
potential, the energy-momentum
tensor vanishes at all time as conjectured\cite{soliton,sft}.\footnote{For
early studies of open string tachyon dynamics, see \cite{old}.}

Given these results,
we can ask: how does the energy momentum tensor (and other
closed string
sources) evolve with time for the rolling tachyon solution in
the presence
of electric flux?
In this paper we answer this question by constructing a two parameter
family of time dependent solutions, labelled by the initial electric field
$e$ and the
initial value $\tl$ of the tachyon. We find,
first of all, that the solution carries fundamental string
charge as expected. At arbitrary values of $\tl$ and  the electric
field $e$ (below the critical limit), the energy-momentum tensor
$T_{\mu\nu}$ of the solution splits into a sum of
contribution from two sources, -- the rolling tachyon, and a
uniform density of fundamental strings\cite{Dab}, localised on the
initial location of the brane, and stretched along the direction
of the electric field. The contribution to $T_{\mu\nu}$ from the
fundamental string charge remains constant in time, whereas the
rolling tachyon contribution has the same form as that in the
absence of electric field, except for a change in the overall
normalisation and a time dilation which depends on the strength of
the electric field. Since the time evolution of the rolling
tachyon contribution gets affected by the value of the electric
field, it shows that the two systems are coupled together. In a
suitable limit where the initial value of the tachyon approaches
the minimum of the potential ($\tl \to 1/2$ ) and the background
electric field goes to its critical value ($e\to 1$), the
contribution from the rolling tachyon drops out, leaving behind
only the time independent fundamental string
contribution\footnote{This implies that the electric flux
solution obtained this way is a stationary solution.}. This
coincides with the source terms discussed in \cite{Dab}. However
our analysis also shows that this source has divergent components
for the higher massive closed string states. The physical
significance of this divergence has not been investigated in this
work. However, since fundamental string configurations are valid
configurations in string theory, our results can be turned around to conclude
that divergent higher level contribution to a boundary
state does not necessarily imply a singularity of the configuration that
it describes.
(Different aspects of
higher level contribution to the boundary state associated with rolling
tachyon configuration have been discussed in \cite{private}.)

We carry out our analysis in two different ways. In the first
approach discussed in section \ref{s2} we assume the existence of
a space-time effective field theory for the tachyon, possibly
containing infinite number of higher derivative terms, and use the
results of \cite{9908142} to write the sources for the massless
closed string fields in presence of the background electric field
in terms of the sources in absence of the background electric
field. This does not require knowledge of the explicit form of the
effective action, and is done by the following steps:

\begin{enumerate}

\item
Using the results of \cite{9908142} we first rewrite the action
at a
nonzero background electric field $e$ using the open string metric and the
non-commutative $*$ product.
This relates a solution in the presence of background electric field to a
solution in the non-commutative theory with zero background electric
field.
The dependence of the solution on $e$
in the
non-commutative theory comes only through the $*$ product and the open
string
metric.

\item
Now if we look for a solution which depends only on one space-time
direction in the non-commutative theory with zero background
electric field, then one can immediately relate this to a solution
in commutative theory with zero background field, since
non-commutativity does not play any role if the configuration
depends on only one direction. This allows us to relate a solution
in the non-commutative theory, and hence in the commutative theory
with background electric field $e$, to a solution in the
commutative theory with vanishing background electric field.
In the latter theory, the
dependence of the solution on $e$ comes from the dependence of
the open string metric on $e$.
\item The
solution in commutative theory with zero background field and
non-trivial (constant) open string metric can be
further related to the solution in trivial open string metric
background by a linear
coordinate transformation that takes the open string metric to the
trivial metric.

\end{enumerate}

This allows us to construct a solution in the theory with background
electric field and trivial metric in terms of a solution in the theory
with zero background
electric field and trivial metric.
The sources, evaluated at
this pair of solutions
in the two different theories can also be related by using their tensorial
property under this coordinate transformation. Using this and the chain
rule of differentiation one can then write down the sources derived from
the Lagrangian with a non-zero background electric field and trivial
metric in terms of the
sources
derived from the Lagrangian with zero background electric
field and trivial metric.
Since the sources in absence of the background field are already known
\cite{0203265}, this gives the sources in the presence of the background
electric field.

Our second approach discussed in section \ref{s3} is to directly
analyse the BCFT corresponding to the rolling tachyon solution
coupled to the background electric field. We construct this BCFT by adding
appropriate perturbation to the BCFT
describing the unstable D-brane with a constant background
electric field. In \cite{0203211},
the perturbation associated with the rolling tachyon solution was
identified from the spatially homogeneous solution to the
linearised equation of motion for the tachyon in string field
theory. Proceeding along the same line, we first find the
solution to the linearised equation of motion in presence of the
background electric field, and use it to identify the rolling tachyon
perturbation in presence of
background field. Having identified the relevant BCFT we analyse
it following the approach of \cite{0202210,0203211}, namely Wick-rotate
the time direction to go to a theory with all spatial directions.
The resulting theory without the perturbation term is equivalent to the
standard magnetic BCFT analysed in
\cite{fradkin,abouel,9503014,9908142} with an imaginary magnetic field. We
do all the
explicit analysis with a real magnetic field and at the end
recover results in the original electric theory by performing
the inverse Wick-rotation and an analytic continuation which makes
the magnetic field imaginary. We study the perturbation
describing the
rolling tachyon in this magnetic BCFT and argue, using
the idea of {\it locality} of boundary operators introduced in
\cite{9811237}, that this deformation is exactly marginal.
This gives a two parameter family of BCFT's labelled by the background
electric field $e$ and the strength $\tl$ of the deformation.
We then
construct the boundary state associated with these BCFT's
by generalising the
construction given in
\cite{9503014,9402113,9811237,0108238} to the present case. Once the
boundary
state is known, we can extract the massless closed string sources
from this boundary state\cite{9604091,9707068,9912161}. These reproduce
the
results obtained
through the target space analysis of section \ref{s2}. The new
information that one gets from the boundary state analysis is the
divergence of the boundary state for the higher level terms in the
limit $\tl\to 1/2$, $e\to 1$ discussed earlier. The divergence is
demonstrated by performing an explicit computation of the terms at
the next higher level.

All the above discussions have been made in the case of bosonic
string theory. We generalise the analysis to the superstrings in
section \ref{superstrings}. We have computed only the massless
sources by generalising the analysis of \cite{0203265}. Although
we have not computed the full boundary state or performed any
computation at the next higher level, one might expect to see
similar divergences in the $\tl\to 1/2$, $e\to 1$ limit in this
case also.

In section \ref{s4} we discuss a candidate low energy world-volume
effective
action following \cite{9909062,effective,0003221,0203265,0204143}
which reproduces the results for the sources of massless closed
string fields. We also discuss its coupling to the supergravity
fields including the massless RR backgrounds.

\sectiono{Target Space Analysis} \label{s2}

We denote by $\SSS(T; g_{\mu\nu}, b_{\mu\nu})$ the effective
action describing the dynamics of the tachyon field on
a D$p$-brane in the presence of constant background closed string metric
$g_{\mu\nu}$ and anti-symmetric tensor field
$b_{\mu\nu}$. By the result of \cite{9908142}, this
effective action is equivalent to another action:\footnote{We have
absorbed factors of $2\pi$ into the definition of $b_{\mu\nu}$ and
$\Theta_{\mu\nu}$ compared to \cite{9908142} in order to simplify
various formul\ae.}
 \be \label{e1a} \SSS(T; g_{\mu\nu},
b_{\mu\nu}) = 
\sqrt{\det(g+b)\over \det G}\, \SSS_\Theta (T; G_{\mu\nu},
b_{\mu\nu}=0), \ee
 where $\SSS_\Theta$ is related to $\SSS$ by the
replacement of all products by $*$-products with non-commutativity
parameter $\Theta$, and $G_{\mu\nu}$ is the open string metric.
The inverse $G^{\mu\nu}$ of $G_{\mu\nu}$ and $\Theta^{\mu\nu}$ are
given in terms of $g_{\mu\nu}$ and $b_{\mu\nu}$ by the relations:
\be \label{e1} G^{\mu\nu} = [(g+b)^{-1}]_S^{\mu\nu}, \qquad
\Theta^{\mu\nu} = [(g+b)^{-1}]_A^{\mu\nu}, \ee where the
subscripts $S$ and $A$ denote the symmetric and anti-symmetric
parts of the corresponding matrices respectively. In general the
fields $T$ appearing on the two sides of \refb{e1a} are related by
a field redefinition, but for the configurations we shall be
considering where the tachyon depends only on the time coordinate,
this field redefinition is identity. 

The equivalence \refb{e1a} implies that if we can construct a classical
solution of the equations of motion of $\SSS_\Theta (T; G_{\mu\nu},
b_{\mu\nu}=0)$, then we also have a solution of the equations of motion of
$\SSS(T; g_{\mu\nu}, b_{\mu\nu})$. We shall focus on solutions which 
depend
on only the time coordinate; and in this case we can replace the
$*$-product by ordinary product. Thus we need to find solutions of the
equations of motion of $\SSS(T; G_{\mu\nu},
b_{\mu\nu}=0)$. For constant $G_{\mu\nu}$, these solutions, in turn, can
be computed from solutions of the
equations of motion of $\SSS(T; \eta_{\mu\nu}, b_{\mu\nu}=0)$ by a linear 
transformation on the coordinates
that converts the metric $\eta_{\mu\nu}$ to $G_{\mu\nu}$.\footnote{Related
technique has been used earlier in \cite{0007226}.}

To be more specific, we shall consider a background of the form:
\be \label{e2}
g_{\mu\nu}=\eta_{\mu\nu}, \qquad b_{01}=-b_{10} = e\, ,
\ee
with all other components of $b_{\mu\nu}$ being zero.
Since a background $b_{01}$ is equivalent to switching on an electric
field on the D-brane world-volume, throughout this paper we shall refer to
such background as background electric field.
Eqs. \refb{e1} and \refb{e2} give:
\be \label{e3}
G_{00} = -(1-e^2), \qquad G_{11} = (1-e^2), \qquad G_{ij}=\delta_{ij}
\quad \hbox{for} \quad i,j \ge 2, \qquad \Theta^{01}=e/(1-e^2)\, ,
\ee
with all other components of $G_{\mu\nu}$ and $\Theta^{\mu\nu}$ being
zero. $G_{\mu\nu}$ can be converted to $\eta_{\mu\nu}$ by a rescaling
of
$x^0$ and $x^1$ by $\sqrt{1-e^2}$. Thus if $T=F(x^0)$ is a solution of the
equations of motion of $\SSS(T; \eta_{\mu\nu},
b_{\mu\nu}=0)$, then $T=F(\sqrt{1-e^2}\, x^0)$ will be a solution of the
equations of motion of $\SSS(T; G_{\mu\nu},
b_{\mu\nu}=0)$ and also of $\SSS_\Theta
(T; G_{\mu\nu},
b_{\mu\nu}=0)$. By the equivalence \refb{e1a}, it
is then
also a solution
of the equations of motion of $\SSS(T; g_{\mu\nu}, b_{\mu\nu})$.

Thus if we knew the solution $F(x^0)$, we could find the solution
of the equations of motion of $\SSS(T; g_{\mu\nu}, b_{\mu\nu})$. In
actual practice, however, we do not know the solution $F(x^0)$,
but only know the energy momentum tensor associated with this
solution\cite{0203211,0203265}. Let us denote this by 
$T^{(0)}_{\mu\nu}(x^0)$: \be
\label{e4} T^{(0)}_{\mu\nu}(x^0) = -2 \, {\delta \SSS(T=F(x^0);
G_{\mu\nu}, b_{\mu\nu}=0) \over \delta
G^{\mu\nu}(x)}\Bigg|_{G_{\mu\nu}=\eta_{\mu\nu}}\, . \ee
(Throughout this paper we shall omit writing the dependence of
various closed string sources on coordinates transverse to the
D-brane, which is just a delta function of the transverse
coordinates.) Then the energy momentum tensor \ben \label{e5} \wt
T_{\mu\nu}(x^0) &=& -2 \, (-\det G)^{-1/2} \, {\delta
\SSS(T=F(\sqrt{1-e^2}\, x^0); G_{\mu\nu}, b_{\mu\nu}=0) \over
\delta G^{\mu\nu}(x)}\, , \een for $G_{\mu\nu}$ given in
\refb{e3}, is related to $T^{(0)}_{\mu\nu}(x^0)$ by the standard
transformation laws under $(x^0, x^1) \to \sqrt{1-e^2} \, (x^0,
x^1)$. This gives \ben \label{e6} && \wt T_{ab}(x^0) = (1-e^2) \,
T^{(0)}_{ab}(\sqrt{1-e^2}\, x^0), \qquad \wt T_{ai} (x^0) =
\sqrt{1-e^2} \, T^{(0)}_{ai}(\sqrt{1-e^2}\, x^0), \nonumber \\
&&
\wt T_{ij}(x^0) =T^{(0)}_{ij}(\sqrt{1-e^2}\, x^0),
\qquad
\hbox{for} \quad a,b=0,1, \quad i,j\ge 2\, .
\een
For the case at hand, we know the explicit form of $T^{(0)}_{\mu\nu}$ from
the analysis of \cite{0203211,0203265}. For definiteness let us consider
the case of bosonic string theory. In this case:
\be \label{e7}
T^{(0)}_{00} = {1\over 2} \, \TT_p \, (1 + \cos(2\pi\tl)), \qquad
T^{(0)}_{r0} = 0, \qquad T^{(0)}_{rs} = 
- \TT_p \, f(x^0) \, \delta_{rs}\,,
\qquad \hbox{for}\quad r,s\ge 1\, .
\ee
Here $\TT_p$ is the tension of the D-$p$-brane, $\tl$ is a parameter
labelling the total energy of the system, and
\be \label{e8}
f(x^0) = {1\over 1 + \sin(\pi\tl) e^{x^0}} + {1\over 1 + \sin(\pi\tl)
e^{-x^0}} -1\, .
\ee
Using eqs.\refb{e6}, \refb{e7} we get
\ben \label{e9}
&& \wt T_{00} = (1-e^2) \, {1\over 2} \, \TT_p \, (1 + \cos(2\pi\tl)),
\qquad
\wt T_{11} = - (1-e^2) \, \TT_p \, f(\sqrt{1-e^2}\, x^0), \nonumber \\
&& \wt T_{ij} = - \TT_p \, f(\sqrt{1-e^2}\, x^0) \, \delta_{ij},
\qquad \hbox{for} \quad i,j\ge 2\, . \een
 The quantities of interest to us are the energy momentum tensor
 and the source for the antisymmetric tensor field
 computed from the action $\SSS(T; g_{\mu\nu}, b_{\mu\nu})$:
 \be \label{e10}
T^{\mu\nu}(x) = 2 {\delta \SSS(T; g_{\mu\nu}, b_{\mu\nu}) \over
\delta g_{\mu\nu}(x)}\bigg|_{g_{\mu\nu}=\eta_{\mu\nu}} , \qquad
S^{\mu\nu}(x) = 2 {\delta \SSS(T; g_{\mu\nu}, b_{\mu\nu}) \over
\delta b_{\mu\nu}(x)}\bigg|_{g_{\mu\nu}=\eta_{\mu\nu}}\, . \ee
 Thus we have
 \be \label{edels}
 \delta \SSS(T; g_{\mu\nu}, b_{\mu\nu}) = {1\over 2} \,\int d^{p+1} x \, 
 \left[T^{\mu\nu}(x)\delta g_{\mu\nu}(x) +S^{\mu\nu}(x)\delta
 b_{\mu\nu}(x)\right].
 \ee
 We shall now consider space-time independent variations $\delta
 g_{\mu\nu}$ and $\delta
 b_{\mu\nu}$. In this case,
 eq.\refb{e1a} gives
\ben \label{e11}
&& \delta \SSS(T; g_{\mu\nu}, b_{\mu\nu}) \nonumber \\
 &=&  \left[
\delta\left(\sqrt{\det(g+b)\over \det G}\right) \, \SSS_\Theta (T;
G_{\mu\nu}, b_{\mu\nu}=0) + \sqrt{\det(g+b)\over \det G}\, \delta
\SSS_\Theta (T; G_{\mu\nu}, b_{\mu\nu}=0) \right] \nonumber \\
&=&  \left[ \delta\left(\sqrt{\det(g+b)\over \det G}\right) \,
\SSS(T; G_{\mu\nu}, b_{\mu\nu}=0) + \sqrt{\det(g+b)\over \det G}\,
\delta \SSS(T; G_{\mu\nu}, b_{\mu\nu}=0) \right]
\, ,\nonumber \\
\een
 where in the last line we have made use of the fact that for $T$
 dependent on only the $x^0$ coordinate we can ignore the $\Theta$
 dependence of the action. 
Eq.\refb{e5} gives:
 \be \label{exz1}
\delta\SSS(T; G_{\mu\nu}, b_{\mu\nu}=0) = -{1\over 2} \, \int d^{p+1} x \, 
\wt T_{\mu\nu}(x) \delta G^{\mu\nu}\, ,
\ee
where, using \refb{e1} we have, 
 \ben \label{e12}
\delta G^{00} &=& -(1-e^2)^{-2} (\delta g_{00} + 2 e \delta b_{01} - e^2
\delta g_{11})\, , \nonumber \\
\delta G^{11} &=& -(1-e^2)^{-2} (\delta g_{11} - 2 e \delta b_{01} - e^2
\delta g_{00})\, , \nonumber \\
\delta G^{ij} &=& -\delta g_{ij}, \qquad \hbox{for} \quad i,j\ge 2\, ,
\nonumber \\
\delta\left(\sqrt{\det(g+b)\over \det G}\right) &=& {1\over 2}
(1-e^2)^{-3/2} (e^2 \delta g_{00} - e^2 \delta g_{11} + 2 e \delta
b_{01} )\, . \een
In order to evaluate the right
 hand side of \refb{e11}, we also need to compute $\SSS(T; G_{\mu\nu},
 b_{\mu\nu}=0)$ for the solution $T=F(\sqrt{1-e^2} x^0)$. This is done as 
follows.
 Let
us consider a space-time independent variation $\delta G_{22}$ of
$G_{22}$ with all other $\delta G_{\mu\nu}=0$. In this case, for
the background of the form considered here, the only contribution
to $\delta\SSS(T; G_{\mu\nu}, b_{\mu\nu}=0)$ will come from the
overall multiplicative factor of $\sqrt{-\det G}$ in the Lagrangian 
density. Thus we have: \be \label{e9a} \delta
\SSS(T; G_{\mu\nu}, b_{\mu\nu}=0) = {1\over 2} \, \delta G_{22} \,
\SSS (T; G_{\mu\nu}, b_{\mu\nu}=0) \, , \ee since $G^{22}=1$.
 On other hand eq.\refb{e5} gives
  \be
\label{e9aa} \delta \SSS (T; G_{\mu\nu}, b_{\mu\nu}=0) = -{1\over
2} \,  \int d^{p+1} x \, (-\det G)^{1/2} \, \delta G^{22}\,   \wt
T_{22} = {1\over 2} \, \int d^{p+1} x \, (-\det G)^{1/2} \, \delta
G_{22} \, \wt T_{22} \, , \ee since $\delta G_{22} = -\delta
G^{22}$. Comparing \refb{e9a} and \refb{e9aa} we get
  \be \label{e9b}
   \SSS(T; G_{\mu\nu}, b_{\mu\nu}=0) =\int d^{p+1}
x \, (-\det G)^{1/2} \, \wt T_{22}\, . \ee
 Eqs.\refb{edels}-\refb{e12}, and  \refb{e9b} now 
give:
 \ben \label{e13} && \delta \SSS(T;
g_{\mu\nu}, b_{\mu\nu}) \equiv {1\over 2} \, \int d^{p+1} x  \,
[T^{\mu\nu} \delta
g_{\mu\nu} + S^{\mu\nu} \delta b_{\mu\nu} ]\nonumber \\
&=&
{1\over 2}\int d^{p+1} x \, \bigg[ \delta g_{00} \{  (1-e^2)^{-3/2} \wt
T_{00} - e^2
(1-e^2)^{-3/2} \wt T_{11} + e^2
(1-e^2)^{-1/2} \wt T_{22} \} \nonumber \\
&& \qquad + \delta g_{11} \{  (1-e^2)^{-3/2} \wt T_{11} - e^2
(1-e^2)^{-3/2} \wt T_{00} - e^2
(1-e^2)^{-1/2} \wt T_{22} \} \nonumber \\
&& \qquad + 2 e \delta b_{01} \, \{ (1-e^2)^{-3/2} \wt T_{00} -
(1-e^2)^{-3/2}
\wt
T_{11} +    (1-e^2)^{-1/2} \wt T_{22}\} \nonumber \\
&& \qquad + (1-e^2)^{1/2} \wt T_{ij} \delta g_{ij}
\bigg] \, . \nonumber \\
\een
Since $\delta g_{\mu\nu}$ and $\delta b_{\mu\nu}$ are arbitrary constants,
from
\refb{e13} we can compute space-time
integrals of $T_{\mu\nu}$ and $S_{\mu\nu}$. If we assume that the relation
holds also for the integrands, then we get
\ben \label{e14}
T_{00}&=&T^{00} =  \bigg[{1\over 2} \, e^2 \, (1-e^2)^{-1/2} \TT_p \, (1 +
\cos(2\pi\tl))\bigg] + \bigg\{{1\over 2} \,(1-e^2)^{1/2} \TT_p \, (1 +
\cos(2\pi\tl))\bigg\} \, ,
\nonumber \\
T_{11} &=& T^{11} = -\bigg[ {1\over 2} \, e^2 \, (1-e^2)^{-1/2} \TT_p \,
(1 +
\cos(2\pi\tl)) \bigg]
- \bigg\{ (1-e^2)^{1/2} \TT_p f(\sqrt{1-e^2}\, x^0) \bigg\} \, , \nonumber
\\
T_{ij} &=& T^{ij} = -\bigg\{(1-e^2)^{1/2}  \TT_p f(\sqrt{1-e^2}\,
x^0)\delta_{ij} \bigg\}
\, ,
\nonumber \\
S_{01} &=& -S^{01} = - \bigg[{1\over 2}\, e \, (1-e^2)^{-1/2} \TT_p \, (1
+
\cos(2\pi\tl))\bigg] \, .
\een
Note that in defining $S_{01}$ we have taken into account the fact that
$\delta\SSS$ receives contribution from ${1\over 2}S^{01}\delta b_{01}$ as
well as
${1\over 2}S^{10}\delta b_{10}$.

We now need to examine to what extent it is justified to equate the
integrands in \refb{e13} to arrive at \refb{e14}. First of all, note that
the dependence of $T_{\mu\nu}$ and $S_{\mu\nu}$ on the
spatial coordinates is trivial (independent of the tangential directions,
and proportional to a delta function involving the transverse coordinates
which is
understood in \refb{e14}). Thus the only question is if we can equate the
integrands under the $x^0$ integral. Since 
$T_{0\mu}$ and
$S_{0\mu}$ are time independent due to their conservation laws, the
answers for these quantities given in \refb{e14} are certainly valid. As
for the other
quantities, we note that the detailed time dependence of these quantities
depends to a large extent on the precise off-shell definition of the
metric since we need to compute the change in the action under a local 
variation of the metric and anti-symmetric tensor field. As a result, 
these quantities can be changed by changing the definition of the 
off-shell continuation of the metric {\it e.g.} by redefining the 
off-shell vertex operator by a conformal transformation. Thus the 
expressions given
in
\refb{e14} are equally good choices to any other expressions consistent
with the integrated equation \refb{e13}.
Nevertheless,
we shall see in later sections that the boundary state analysis, which 
comes with a precise convention for coupling of off-shell closed string 
states to a D-brane, leads
to the same expressions for the time evolution of the various
sources as given in \refb{e14}. 

$T_{\mu\nu}$ and $S_{\mu\nu}$ given in \refb{e14} can be interpreted as a
sum of
the contribution from a configuration of fundamental strings (shown in
square bracket) and from rolling tachyon in the absence of electric field
(shown in curly brackets). Note however that the evolution of the
contribution from rolling tachyon slows down by a factor of $\sqrt{1-e^2}$
in the presence of fundamental string charge. Thus the two systems do not
decouple. We can get pure fundamental string background without any
contribution from rolling tachyon by taking the limit
\be \label{elimit}
e\to 1, \qquad \tl\to {1\over 2}, 
\qquad (1-e^2)^{-1/2} (1 +\cos(2\pi\tl))  \quad
\hbox{fixed}\, .
\ee
However, as shown in appendix \ref{divergence}, the
contribution to the
boundary state from higher level closed string states blows up in this
limit.
The meaning of this divergence is not entirely clear to us. 

These results can be generalised to the case of decay
of unstable D-$p$-branes and brane-antibrane systems in superstring
theories. In fact the final formula for $T_{\mu\nu}$ and $S_{\mu\nu}$ are
given by the same expressions as \refb{e14}, all that changes is the
expression for $f(x^0)$. In this case we have\cite{0203265}
\be \label{exx1}
f(x^0) = {1\over 1 + \sin^2(\pi \tl) e^{\sqrt 2 x^0}}
+ {1\over 1 + \sin^2(\pi \tl) e^{-\sqrt 2 x^0}} -1\, .
\ee
Similarly we can generalise the results to the case where the tachyon
begins rolling at the top of the potential with a non-zero velocity so
that the total energy of the system is larger than the tension of the
brane. This requires a replacement of $\cos(2\pi\tl)$ by $\cosh(2\pi\tl)$
and appropriate replacements for $f(x^0)$ as given in \cite{0203265}.

\sectiono{Boundary Conformal Field Theory Analysis} \label{s3}

It was shown in ref.\cite{0203211} that in the absence of
background electric (or $b$) field, the class of time dependent
solutions describing rolling of a D-$p$-brane away from the
maximum of the tachyon potential is given by perturbing the
boundary conformal field theory (BCFT) describing the original
D-brane by the operator
 \be \label{ex1} \tl\, \int dt
\cosh(X^0(t))\, , \ee
 where $\tl$ is a constant
parametrising the initial value of the tachyon, and $t$ denotes
the coordinate labelling the boundary of the world-sheet. This
perturbation was identified in ref.\cite{0203211} from the
spatially homogeneous solution to the linearised equation of
motion for the tachyon in string field theory. Such a deformation
gives rise to a new BCFT since $\cosh(X^0)$ is an exactly marginal
operator, and hence generates a solution of the equations of
motion of open string theory.

In this section we shall generalise this construction to D-branes
in the presence of a background electric field given in eq.
(\ref{e2}). As discussed in the previous section, the space
independent solution for the tachyon in presence of this
non-trivial background can simply be obtained by incorporating the
coordinate transformation $(x^0, x^1) \to \sqrt{1-e^2} \, (x^0,
x^1)$ into the solution in absence of any background field. This,
in turn means that  the
rolling tachyon solution in the presence of background field
configuration $b_{01}=e$, is generated by the following
deformation,
 \be \label{enewm1} \tl\, \int dt \cosh(\sqrt{1-e^2}
\, X^0(t))\, .
 \ee
  In the following we shall show that  the deformation (\ref{enewm1})
is exactly marginal, and analyse the deformed
BCFT. We shall also find the boundary state
associated with this BCFT.

\subsection{The Boundary Conformal Field Theory} \label{bcft}

In this subsection we shall demonstrate that \refb{enewm1}
generates an exactly marginal deformation of the BCFT describing
the D-brane in the background \refb{e2}. In order to do so we
shall show that this represents an exactly marginal deformation in
the Wick rotated theory obtained by the replacement $X^0 \to -i
X^0$. Under this rotation the closed string metric $g_{\mu\nu}$
and the antisymmetric tensor field $b_{\mu\nu}$ go to
$\bg_{\mu\nu}$ and $\bbb_{\mu\nu}$ respectively, given by,
 \be
\label{eas1} \bg_{\mu\nu} = \delta_{\mu\nu}, \qquad \bbb_{01} =
-\bbb_{10} = -ie \equiv \bar e\, , \ee
 with other components of $\bbb_{\mu\nu}$
being zero. After performing Wick rotation on the world-sheet as
well, one gets the following boundary condition:
 \be \Big(\bg_{\mu \nu}~ \partial_{n} X^{\nu} ~+~ i \bbb_{\mu \nu}
~\partial_{t} X^{\nu} \Big)\Big |_{\partial \Sigma} = 0, \label{openbc}
\ee
 where $\partial_n$ and $\partial_t$ denote respectively the
normal and tangential derivatives at the boundary $\partial
\Sigma$. For simplicity we have restricted our analysis to the
case of D-25-brane, but the generalisation to an arbitrary
D-$p$-brane is straightforward. The boundary conformal field theory at 
hand
is same as the one \cite{9908142} corresponding to an imaginary
$B$-field $\bbb_{\mu\nu}$ switched on along spatial Neumann
directions. In the following we shall consider the case of a real
$\bbb$-field, {\it i.e.} a real magnetic field $\bar e$ in the 0-1
plane, and present all the relevant formul{\ae} in that context.
Any given formula can be translated to the one corresponding to an
electric field by the analytic continuation, \be \bar e
\rightarrow  -i e, \label{analytic} \ee together with an inverse
Wick rotation $X^0\to i X^0$.

The expressions for the effective Euclidean open string
metric $\bG$ and the non-commutativity parameter $\bT$ are given by,
\be
\bG^{\mu \nu}  = \left[(\bg +\bbb)^{-1}\right]_S^{\mu \nu}, 
\qquad
\bT^{\mu \nu} = \left[(\bg +\bbb )^{-1}\right]_A^{\mu \nu}. 
\label{Gtheta}
\ee
It will be convenient for our discussion to
introduce the vielbeins of the open and
closed string metrics and the
corresponding local coordinates. These are defined
by the following equations:
\ben
\begin{array}{cc}
\bG_{\mu \nu}= V^a_{~\mu}V^a_{~\nu} = (V^TV)_{\mu \nu}, &
 \bg_{\mu \nu}= v^a_{~\mu}v^a_{~\nu} = (v^Tv)_{\mu \nu}, \\
&\\ Z^a=V^a_{~\mu} X^\mu, &  Y^a=v^a_{~\mu} X^\mu \,,
\end{array}
\label{GgZY}
\een
where $a, ~\mu = 0, \cdots, 25$ and the matrix
notation for various quantities has the obvious meaning.
For the case under study, we have
\be
\bg=\one_{26}, \qquad
\bbb= \pmatrix{0 & \bar e & 0 \cr -\bar e & 0 & 0 \cr 0 & 0 & \zero_{24}},
\label{gB}
\ee
\be
\bG=\pmatrix{\displaystyle{1+\bar e^2}  & 0 & 0 \cr 0 &
\displaystyle{1+\bar e^2}
 & 0 \cr 0 & 0 & \one_{24}}, \qquad
\bT= \pmatrix{0 & -\displaystyle{\bar e \over 1+\bar e^2} & 0 \cr
\displaystyle{\bar e \over 1+\bar e^2} & 0 & 0 \cr 0 & 0 & \zero_{24}},
\ee
and we shall choose
\be
V = \pmatrix{\displaystyle\sqrt{1+\bar e^2}  & 0 & 0 \cr 0 &
\displaystyle \sqrt{1+\bar e^2}
 & 0 \cr 0 & 0 & \one_{24}}, \qquad
v = \pmatrix{\displaystyle{1\over \sqrt{1+\bar e^2}} &
\displaystyle {\bar e \over \sqrt{1+\bar e^2}} & 0 \cr
- \displaystyle {\bar e \over \sqrt{1+\bar e^2}} &
\displaystyle {1\over \sqrt{1+\bar e^2}} & 0 \cr 0 & 0 & \one_{24}} .
\label{Vv}
\ee
The relation between the two local
frames is given by,
\be
Z=LY, \qquad L=Vv^{-1}=\pmatrix{1 & - \bar e & 0\cr \bar e & 1 & 0\cr 0 &
0 & \one_{24}}\, .
\label{ZY}
\ee

The boundary two point functions \cite{fradkin,abouel,9908142} of the 
local
coordinates $Z^a$ take the form ($\alpha^{\prime}=1$),
\be
\label{ezcor} \la Z^a(x) Z^b(y) \ra = - \delta^{ab} \log (x-y)^2
~+~ {i ~ \pi} ~ \bT_Z^{ab} \epsilon (x-y), \label{ZZ}
\ee
where,
\be
\bT_Z = V \bT V^T,
\ee
and $\eps(x)$ is the sign function. Let
us now define the following boundary operators,
\be
J^1_{Z^a}(x)=\cos\left(Z^a(x)\right), \quad
J^2_{Z^a}(x)=\sin\left(Z^a(x)\right), \quad J^3_{Z^a}(x)={i \over
2} \partial Z^a(x). \label{currents}
\ee
Using the two point
function (\ref{ZZ}) it is straightforward to show that these
operators have conformal dimension 1. We shall now argue along the
line of ref.\cite{9811237} that the operator \be \label{eas3} \int
dt \, {\cal O}^{Z^0}_{\tl}(t) \equiv \tl \int dt \, J^1_{Z^0}(t) =
\tl \int dx \, \cos\Big(\sqrt{1+\bar e^2} X^0(t) \Big) \, ,
\label{calO} \ee describes an exactly marginal deformation. It was
shown in \cite{9811237} that \refb{eas3} describes an exactly
marginal deformation if ${\cal O}^{Z^0}_{\tl}$ is
{\it self-local}, {\it
i.e.} in a correlation function involving ${\cal O}^{Z^0}_{\tl}(x)
{\cal O}^{Z^0}_{\tl}(y)$, the result for $y>x$ is related to that
for $y<x$ by an analytic continuation in the complex $y$ plane,
and the result of the analytic continuation does not depend on
whether it is done in the upper half plane or the lower half
plane. Using the two point function
\refb{ezcor} it is straightforward to verify that
this condition is satisfied.\footnote{This argument can be easily
extended to the
case of the arbitrary linear combination
${\cal O}^{Z^a}_{\tl_1, \tl_2} = \tl_1 J^1_{Z^a}+\tl_2
J^2_{Z^a}$. In fact using the notion of {\it mutual locality} between
two operators introduced in \cite{9811237} one can show that there
exist special values of the magnetic field, namely $\bar e=n, n
\in {\bf Z}$ for which ${\cal O}^{Z^a}_{\tl_1, \tl_2}$ and ${\cal
O}^{Z^b}_{\tl_1, \tl_2}$, for $a\neq b$, become mutually local,
and hence any linear combination of them represents a marginal
deformation. These correspond to the special conformal points
(magic) found in \cite{9503014} where two cosine perturbations
were simultaneously switched on along orthogonal directions in
presence of a magnetic field. These special conformal points do
not occur in the electric theory due to the fact that the
non-commutativity parameter is imaginary in this case.}

This establishes the exact marginality of the operator
$\cos(\sqrt{1+\bar e^2} X^0)$.
Making the analytic continuation $\bar e\to -ie$ and the inverse
Wick-rotation $X^0\to i X^0$ we arrive at the original
BCFT with a time-like direction
where the above marginal operator becomes
$\cosh(\sqrt{1-e^2} X^0)$. Thus we
can perturb the original theory by
$\tl\int dt \cosh(\sqrt{1-e^2} X^0(t))$ to generate
rolling tachyon solutions in the presence of electric field.

In order to find the energy-momentum tensor associated with this rolling
tachyon solution, we need to evaluate the boundary state associated with
the deformed BCFT.
This can be done by first working with the Euclidean theory with
real magnetic
field $\bar e$ and then making the replacement $X^0 \to i X^0$, $\bar e
\to -i e$.
We shall do this next and show that
the result agrees with the ones found in section \ref{s2}.
The construction in the Euclidean theory will proceed according to the
following steps:
\begin{enumerate}

\item First we shall compactify the theory in a specific manner, and
construct the boundary state $|\BB_\bbb\ra$ corresponding to the
unperturbed BCFT ($\tl =0$) including the background magnetic
field with this specific compactification.

\item Then we study the effect of switching on the perturbation $\tl$ on
the
boundary state in the compactified theory, and compute the deformed
boundary state $|\BB_{\bbb,\tl}\ra$.

\item Finally we take the decompactification limit by removing all the
winding modes
in the expression for the boundary state. This gives the deformed boundary
state $|\BB^{\infty}_{\bbb,\tl}\ra$ in the non-compact theory.

\end{enumerate}

\subsection{Compactification and Enhancement of $SU(2)$ Symmetries}
\label{compactification}

Following ref.\cite{9402113,9811237} we shall first construct the boundary
state in a theory with suitable compactification and then go to
its universal cover to get the answer in the non-compact limit. In
this subsection, therefore we shall discuss some issues involving
compactifications. We want to construct the boundary state
corresponding to the marginal deformation given by the operator
\refb{eas3}. Therefore the compactifications of our interest are
the ones that maintain the periodicity of this operator. At the
same time, we would like to ensure that in the compactified theory
there is an enhanced $SU(2)_L\times SU(2)_R$ symmetry in the
closed string sector which can be used to organise the boundary
state as in \cite{9402113,9811237}. Naively one might think that
periodicity of $\cos(Z^0)$ would require compactifying $Z^0$ on a
circle of unit radius keeping $Z^1$ non-compact, {\it i.e.} making
the identification $(Z^0, Z^1) \equiv (Z^0+2\pi, Z^1)$. However in
this case the radius of the compact circle, measured in the closed
string metric, is $1/\sqrt{1+ \bar e^2}$, and hence we do not get
enhanced $SU(2)_L\times SU(2)_R$ gauge symmetry in the closed
string sector. The compactification which achieves this is \be
\label{eas4} (Y^0, Y^1) \equiv (Y^0 + 2\pi, Y^1)\, . \ee {}From
\refb{ZY} we have
\begin{eqnarray}
& Z^0= Y^0 - \bar e Y^1, \qquad Z^1= \bar e Y^0 + Y^1, \qquad Z^\mu=Y^\mu
\quad
\hbox{for} \, \, \mu\ge 2 \nonumber \\
\to & \qquad Y^0= \displaystyle {Z^0 \over 1+\bar e^2} +
\displaystyle {\bar e Z^1 \over 1+\bar e^2} ,
\qquad Y^1= -\displaystyle {\bar e Z^0 \over 1+\bar e^2} +
\displaystyle {Z^1 \over 1+\bar e^2}.
\label{ZYYZ}
\end{eqnarray}
This, together with \refb{eas4} gives
\be \label{eas5}
(Z^0, Z^1) \equiv (Z^0+2\pi, Z^1+2\pi \bar e)\, .
\ee
The operator \refb{eas3} is clearly invariant under this transformation.
On the other hand since in the $Y^\mu$ coordinate system the closed string
metric is identity, compactifying $Y^0$
with period $2\pi$ implies that the radius of the circle, measured in
closed string metric, is 1. Thus the closed string theory now has enhanced
$SU(2)_L\times SU(2)_R$ gauge symmetry which can be used to organise the
boundary
state as in \cite{9402113,9811237}. The left moving currents are given by,
\be
J_{Y^0_L}^1(u) = \cos \left( 2 Y^0_L (u) \right), \quad
J_{Y^0_L}^2(u) = \sin \left( 2 Y^0_L (u) \right), \quad
 J_{Y^0_L}^3(u) = i \partial Y^0_L (u). \label{Ji}
\ee
Note that all the $J_{Y^0_L}^i$ are well defined operators
since $Y^0$ is compactified on a circle of self-dual radius.

\subsection{The Boundary State in the Compactified Theory} \label{b-state}

We shall now construct the boundary state
$|\BB_\bbb\ra$ for an Euclidean $\hbox{D}25$-brane with a
magnetic $\bbb$-field turned
on, and $Y^0$ compactified on a circle of unit radius. 
It can be expressed as,
 \be |\BB_\bbb\ra
= |\hbox{Bos} ; \bbb \ra \otimes |\hbox{ghost}\ra, \label{BB} \ee
 where $ |\hbox{Bos} ; \bbb \ra $ and $  |\hbox{ghost}\ra$ represent the  
bosonic and ghost parts of the boundary state respectively. We have
\be \label{eghost}
|\hbox{ghost}\ra =
\exp\left(-\sum_{n \geq 0} (\bar b_{-n} c_{-n} ~+~ b_{-n} \bar
c_{-n}) \right) (c_0 ~+~ \bar c_0) c_1 \bar c_1 |0\ra\, .
\ee
where the ghost oscillators have their
usual definition. Construction of the bosonic part follows
\cite{abouel,9503014}.
The closed string overlap condition on the
cylinder corresponding to the open string boundary condition
(\ref{openbc}) reads,
 \ben (\bg ~\partial_{\tau} X ~+~ i\bbb
~\partial_{\sigma} X) \Big |_{\tau = 0} = 0\, . \label{closedbc}
\een
 This, in turn, gives rise to the following condition on
$|\BB_\bbb\ra $ in terms of the oscillators $\bet_n$'s and $\bar
\bet_n$'s of the coordinates $Y$,
\be \label{eosci} \left[ \bet_n
~+~ M_Y ~\bar \bet_{-n} \right] |\BB_\bbb\ra = 0,
\qquad \forall n\in {\bf Z},
\ee
where the matrix $M_Y$ is given by,
 \ben M_Y=
\pmatrix{\displaystyle {1- \bar e^2 \over 1+ \bar e^2} &
\displaystyle {-2 \bar e \over 1+ \bar e^2} & 0 \cr \displaystyle
{2 \bar e \over 1+ \bar e^2} & \displaystyle {1- \bar e^2 \over 1+
\bar e^2} & 0 \cr 0 & 0 & \one_{24}} \, . \label{eas7} \een
Using \refb{eosci} we get,
 \be \label{enon-zero}
|\hbox{Bos};\bbb \ra = N_\bbb \exp \left[-
\sum_{n=1}^{\infty} {1 \over n} ~\bet^T_{-n} ~M_Y~ \bar \bet_{-n}
\right] |\hbox{Bos};\bbb\ra_0 \, ,
\ee
where 
$|\hbox{Bos};\bbb\ra_0$ is the zero mode 
part of the boundary state.
Since we are considering a D25-brane,
the normalisation constant  is given by\cite{9912161},
 \be N_\bbb = K {\cal T}_{25} \sqrt{\det{(\bg+\bbb)}}\, .
\label{NB} \ee
Here ${\cal T}_{25}$ being the $\hbox{D}25$-brane tension and $K$
is a convention dependent numerical factor independent of $\bar b$. Our
final answer for
the energy momentum tensor will be independent of the choice of $K$.

To construct the zero mode part $|\hbox{Bos};\bbb\ra_0$ let us
first also compactify the coordinate $Y^1$ on a circle of radius
$R$. We shall finally take the $R\to \infty $ limit to get the
desired result. Since all the coordinates except for $Y^0$ and
$Y^1$ are noncompact and satisfy Neumann boundary condition,
$|\hbox{Bos};\bbb\ra_0$ contains only the momentum and winding
mode excitations coming from $Y^0$ and $Y^1$. If $n^{(a)},
w^{(a)} \in {\bf Z}, a=0,1$ are respectively momentum and winding
numbers then the above overlap condition for the zero-modes
restricts the eigenvalues appearing in $|\hbox{Bos};\bbb \ra_0$ in
the following way, \ben \left( {n^{(a)}\over R^{(a)}} +
w^{(a)}R^{(a)} \right) + M^a_{Y~b} \left( {n^{(b)}\over R^{(b)}} -
w^{(b)}R^{(b)} \right) =0. \een Using the explicit matrix value
for $M_Y$ one reduces the above equation to:
 \be w^{(0)}R^{(0)} = {n^{(1)} \over \bar e R^{(1)}}, \qquad
w^{(1)} R^{(1)} = - {n^{(0)} \over \bar e R^{(0)}}, \qquad
R^{(0)}=1, \qquad R^{(1)}=R \label{wn} \, .\ee
 So we see that only two
of the four variables $n^{(a)},~w^{(a)}$ are independent. At a
given finite value of $R$ the restriction on $\bar e$ is given by
the flux quantisation law: $\bar e R = a$, where $a$ is an
integer. This gives $n^{(1)} = a w^{(0)}$, $n^{(0)} = a w^{(1)}$.
Thus the sum can be taken over integer values of $w^{(0)}$,
$w^{(1)}$ and we have, \ben \label{eas8} |\hbox{Bos};\bbb\ra_0 &=&
\sum^{\qquad}_{w^{(0)}, w^{(1)}\in {\bf Z}} \exp\left[i (-\bar e
w^{(1)}R + w^{(0)})y^0_L +i (w^{(1)}R + \bar ew^{(0)}) y^1_L
\right. \cr &&\cr &&\left. +i (-\bar e w^{(1)}R - w^{(0)}) y^0_R
+i (- w^{(1)}R + \bar e w^{(0)}) y^1_R \right] |0\ra \, . \een
The non-compact limit, $R\to \infty$ is obtained by keeping only the
$w^{(1)}=0$ term in the sum over $w^{(1)}$. This gives
\be \label{ezero-mode}
|\hbox{Bos};\bbb\ra_0 = \sum_{m\in {\bf Z}/2} \exp\left[-2im (y^0_L 
-y^0_R)
-2im \bar e (y^1_L +y^1_R ) \right] |0\ra\, .
\ee
Note that in this limit $\bar e$ can take any real value.

We now note from eq.(\ref{eas7}) that $M_Y$ is an orthogonal
matrix of determinant one. Then the oscillators, 
\be \bar \gam_n =
M_Y \bar \bet_n=v^{-1}\bar\alpha_n, \qquad n\neq 0, \label{gamma}
\ee 
where $\alpha^\mu_n$, $\bar\alpha^\mu_n$ denote the oscillators of
$X^\mu$,
obey the same commutation relations as $\bar \bet_n$'s and
therefore as far as construction of basis states is concerned they
are as good as the $\bar \bet_n$ oscillators. Moreover all the
Virasoro oscillators $\bar L_n$'s take the same form in terms of
the $\bar \gam$ oscillators as in terms of the $\bar \bet$
oscillators. Using these facts one can write the state
$|\hbox{Bos};\bbb \ra$ given in (\ref{enon-zero}), \refb{ezero-mode} in 
the 
following
form, \ben |\hbox{Bos};\bbb \ra &=& N_\bbb \exp \left[ -
\sum_{n=1}^{\infty} {1\over n} \left( \bet^0_{-n}\bar \gam^0_{-n}
+ \bet^1_{-n}\bar \gam^1_{-n} +\sum_{i=2}^{25} \alpha^i_{-n}
\bar\alpha^i_{-n}\right) \right] \cr && \times \sum_{m\in {\bf
Z}/2} \exp\left[ -2im(y^0_L -y^0_R) - 2im \bar e y^1 \right] |0\ra
\cr \cr &=& N_\bbb \sum_{j,m} |j,-m,m \ra \ra^{(0)}_{\bar \gam}
\otimes \exp \left[ - \sum_{n=1}^{\infty} {1\over n}
\bet^1_{-n}\bar \gam^1_{-n} - 2im \bar e y^1 \right] |0\ra \otimes
|N\ra_{c=24},\cr && \een where \be |N\ra_{c=24} = \exp
\left[-\sum_{n\ge 1, i\ge 2}{1 \over n} \alpha^i_{-n} \bar
\alpha^i_{-n} \right] |0\ra. \label{N} \ee The state $|j,-m,m \ra
\ra^{(0)}_{\bar \gam} $ is constructed by following the two steps:
construct the  Virasoro Ishibashi state $|j,-m,m \ra \ra^{(0)}$ in
the representation of $SU(2)_{Y^0_L} \otimes SU(2)_{Y^0_R}$
\cite{9402113,9811237}. Then replace the $\bar \bet $ oscillators
by the corresponding $\bar \gam$ oscillators on the right part of
the states appearing in the expansion of $|j,-m,m \ra \ra^{(0)}$.
In fact for $j\neq |m|$, to define the above state one needs to
replace $\bar \bet$ by $\bar \gam$ only in the corresponding
primary state as the whole Ishibashi state can be obtained by
applying various Virasoro oscillators on the primary state. This
automatically gives the $\bar \gam$ dependence of $|j,-m,m \ra
\ra^{(0)}_{\bar \gam} $.

\subsection{Boundary State in the Deformed Theory} \label{sdeformed}

We shall now turn to the boundary state corresponding to the
deformation of the conformal field theory by the boundary
operator,
 \be \int dt \, {\cal O}^{Z^0}_{\tl }(t) = \tl \int dt \,
\cos(Z^0(t))\, . \label{b-operator} \ee
 Using the boundary
condition \refb{eosci}
we can show that on the boundary
$Z^0(t)=2Y^0_L(t)$. This allows us to replace $\cos(Z^0(t))$  in
\refb{b-operator} by $\cos(2 Y^0_L(t))=J^1_{Y^0_L}(t)$. Now, if
$|\BB_{\bbb, \tl} \ra$ denotes the boundary state in the presence
of this perturbation, then, given any closed string vertex
operator $\phi_c$ of ghost number 3, the one point function of
$\phi_c$ on a unit disk in the perturbed BCFT is given by $\la
\BB_{\bbb, \tl}|\phi_c\ra$. Thus we have:\footnote{It is
possible to reduce the integrated boundary deformation in the
action to an exponentiated operator acting on the Hilbert space if
the deformation is exactly marginal. This was shown in
\cite{9811237} very simply by using the self-locality property.
Without using self-locality one can also show this by following
the systematic procedure of renormalisation as was done in
hep-th/9402113 in\cite{9402113}.}
 \be \label{ebexp} \la
\BB_{\bbb, \tl} | \phi_c\ra = \la \BB_{\bbb} | \exp \left[2\pi i
\tl \, Q^1_{Y^0_L} \right] |\phi_c\ra\, , \ee
 where the operators $Q^i_{Y^0_L}$'s
are the $SU(2)_L$ charges in the closed string theory given by,
 \be Q^i_{Y^0_L} = \oint {du \over 2\pi i} ~J_{Y^0_L}^i(u)\, .
\label{Qi} \ee This gives:
 \ben |\BB_{\bbb, \tilde
\lambda}\ra &= & \exp \left[-2\pi i \tl \, Q^1_{Y^0_L} \right]
|\BB_\bbb\ra \nonumber \\ & = &N_\bbb \sum_{j, m} \exp \left[-2\pi
i \tilde \lambda Q^1_{Y^0_L} \right] |j, -m, m\ra \ra_{\bar
\gam}^{(0)} \otimes \exp\left[ -\sum_{n>0} {1 \over n} \bet^1_{-n}
\bar \gam^1_{-n} -2im\bar e y^1 \right] |0\ra \cr && \cr
&& \otimes |N\ra_{c=24} \otimes |\hbox{ghost} \ra, \cr \cr &=& N_\bbb
\sum_{j, m, m'} D^j_{m', -m} |j, m', m\ra \ra_{\bar \gam}^{(0)}
\otimes \exp\left[ -\sum_{n>0} {1 \over n} \bet^1_{-n} \bar
\gam^1_{-n} -2im\bar e y^1 \right] |0\ra \cr && \cr &&
\otimes |N\ra_{c=24} \otimes |\hbox{ghost} \ra, \label{Blambda-bs2} \een
where $D^j_{m', m}$ is simply the spin $j$ representation
matrix of the operator $\exp(-2\pi i\tl Q^1_{Y^0_L})$. 
The values of these matrix elements can be obtained by using the
formula given in ref.\cite{9811237} with the $j={1\over 2}$ 
representation matrix given in
ref.\cite{0203211}.
Notice that
the left parts of all the states appearing in the expansion of
$|j,m',m \ra \ra^{(0)}_{\bar \gam} $ transform in the same and
well defined way under $SU(2)_{Y^0_L}$ while the right parts of
the states do not obey simple transformation rules under
$SU(2)_{Y^0_R}$. But this is sufficient for us to be able to
compute the action of the rotation operator $\exp \left[-2\pi i
\tilde \lambda Q^1_{Y^0_L} \right]$.

Finally we can take the non-compact limit 
of the above
boundary state by simply removing all the winding
sector states from the boundary state \refb{Blambda-bs2}. In this
limit, of the $|j,m',m\ra\ra$ only the $|j,m,m\ra\ra$ states
survive\cite{9402113,9811237,9503014}. Thus we get the final
result,
 \ben |\BB^{\infty}_{\bbb, \tilde \lambda}\ra &= & N_\bbb
\sum_{j, m} D^j_{m, -m} |j, m, m \ra \ra^{(0)}_{\bar \gam} \otimes
\exp\left[ -\sum_{n>0} {1 \over n} \bet^1_{-n} \bar
\gam^1_{-n} -2im \bar e y^1 \right] |0\ra \cr && \cr &&
\otimes |N\ra_{c=24} \otimes  |\hbox{ghost} \ra. \label{Blambda-bs3} \een
 
\subsection{Sources from Boundary State}\label{sources}

In this subsection we shall use the boundary state \refb{Blambda-bs3} to
compute the level one
sources,
namely the energy-momentum tensor $T_{\mu \nu}$
 and the source  $S_{\mu \nu}$ for the antisymmetric
tensor $B$-field.
Since we are interested
in getting the result for the theory with electric
field and the boundary deformation
(\ref{enewm1}), we need to perform a two-fold
operation of inverse
Wick-rotation
$X^0 \to iX^0$ and the analytic continuation (\ref{analytic})
on the quantities we get
directly from the boundary state (\ref{Blambda-bs3}).

Following ref.\cite{0203265} we first notice that the level (1,1)
part of $|\BB^{\infty}_{\bbb, \tilde \lambda}\ra $, 
after the two-fold operation, has the
following general form: \ben \label{egeneral} \int d^{26}k [
(\tilde A_{\mu \nu}(k) + \tilde C_{\mu \nu}(k) ) \alpha^{\mu}_{-1}
\bar \alpha^{\nu}_{-1} + \tilde B(k) (b_{-1} \bar c_{-1} + \bar
b_{-1}c_{-1})] (c_0 +\bar c_0)c_1\bar c_1 |k\ra, \een
where
$\tilde A_{\mu \nu} = \tilde A_{\nu \mu}$ and
$C_{\mu \nu} = -\tilde C_{\nu \mu}$. The conservation law $(Q_B +
\bar Q_B) |\BB_{\bbb, \tilde \lambda}\ra =0$  obtained from this
part of the boundary state reads in the coordinate space, \ben
\partial^{\nu}(A_{\mu \nu}(x)
+ \eta_{\mu \nu} B(x)) =0, \qquad
\partial^{\nu} C_{\mu \nu}(x) = 0\, ,
\een
where $A_{\mu\nu}$, $C_{\mu\nu}$ and $B$ are the Fourier transforms of
$\tilde A_{\mu\nu}$, $\tilde C_{\mu\nu}$ and $\tilde B$ respectively.
This gives the following two conserved
currents,
\ben
T_{\mu \nu}(x) = K_s (A_{\mu \nu}(x) + \eta_{\mu \nu} B(x)),
\qquad
S_{\mu \nu}(x) = K_a  C_{\mu \nu}(x) ,
\label{TS}
\een
where $K_s$ and $K_a$ are two appropriate normalisation constants.

We shall now compute $A_{\mu \nu}(x)$, $C_{\mu \nu}(x)$ and $B(x)$
from $|\BB^{\infty}_{\bbb, \tilde \lambda}\ra$. The non-trivial
part of this calculation is the contribution from the $Y^0$, $Y^1$
parts up to level (1,1). This can be obtained by replacing the
$\alpha^0_{-1}$, $\bar\alpha^0_{-1}$, $\alpha^1_{-1}$, $\bar
\alpha^1_{-1}$ oscillators in the result of \cite{0203211,0203265}
by $-i\bet^0_{-1}$, $-i\bar\gam^0_{-1}$, $\bet^1_{-1}$,
$\bar\gam^1_{-1}$ respectively, and $X^0$ by $-iZ^0=-i(Y^0-\bar e
Y^1)$.\footnote{The factors of $-i$ reflect that we are still in
the Euclidean theory, whereas the results of
\cite{0203211,0203265} were given after inverse Wick rotation.}
The result is proportional to:
\ben \label{eresult}
&&\left[(1-\bet^1_{-1}\bar \gam^1_{-1}) \wh f\left(Y^0(0)-\bar e
Y^1(0)\right) -\bet^0_{-1}\bar\gam^0_{-1} ~\wh g\left(Y^0(0)-\bar e
Y^1(0)\right) \right] |0\ra_{Y^0,Y^1} \nonumber \\ \cr &=&
\left[ (1-\bet^1_{-1}\bar\gam^1_{-1}) \wh f\left(\sqrt{1+\bar e^2}
X^0(0)\right) -\bet^0_{-1}\bar\gam^0_{-1} ~\wh g\left(\sqrt{1+\bar
e^2} X^0(0)\right) \right] |0\ra_{Y^0,Y^1}\, ,
\een
where
\ben \label{edeffbgb}
\wh f(x) &=& \left[ 1 ~+~ \sum_{n=1}^{\infty}
(-\sin(\tilde \lambda \pi) )^n \left( e^{inx} + e^{-inx} \right)
\right] = f(-ix), \cr
\wh g(x)  &=& 1 +\cos(2\pi\tl) - \wh f(x)\,,
\een
with $f(x)$ defined as in eq.(\ref{e8}).

We can now determine $B$, $A_{\mu\nu}$ and $C_{\mu\nu}$ by
comparing \refb{Blambda-bs3} with \refb{egeneral}, and using
\refb{eresult}. We also need to use \refb{gamma} and the fourth
equation in \refb{GgZY} along with the explicit matrix values of
$M_Y$ and $v$ given in eqs.\refb{eas7} and \refb{Vv} respectively,
to translate the $\bet^a_{-1}$ and $\bar\gam^a_{-1}$ into the
oscillators $\alpha^\mu_{-1}$ and $\bar\alpha^\mu_{-1}$ of
$X^\mu$. After the replacement $x^0\to ix^0$, $\bar e\to -ie$, one
gets,
\be B(x) = - K{\cal T}_{25} \sqrt{1-e^2}
f(\sqrt{1-e^2}x^0), \label{B} \ee
\ben A_{00}(x) &=& K
{\cal T}_{25} \left[ (1-e^2)^{-1/2} (1+\cos(2\pi \tilde \lambda ))
- (1-e^2)^{1/2} f(\sqrt{1-e^2}x^0) \right] , \cr
&& \cr
A_{11}(x) &=& -K {\cal T}_{25} \left[ e^2(1-e^2)^{-1/2}
(1+\cos(2\pi \tilde \lambda )) + (1-e^2)^{1/2} f(\sqrt{1-e^2}x^0)
\right], \cr
&&\cr
A_{ij} &=& -K{\cal T}_{25} \,
\delta_{ij} \, \left[ (1-e^2)^{1/2} f(\sqrt{1-e^2}x^0) \right] ,
\quad i, j \geq 2, \cr
&&\cr
C_{01} &=& K{\cal T}_{25}
\left[ e(1-e^2)^{-1/2} (1+\cos(2\pi \tilde \lambda )) \right]\, ,
\label{ASprime} \een with all other components of $A_{\mu\nu}$ and
$C_{\mu\nu}$ being zero.

Using eqs. (\ref{TS}), (\ref{B}) and (\ref{ASprime}) we get
the following non-trivial components for the sources,
\ben
T_{00} &=& K_s K {\cal T}_{25} ~(1-e^2)^{-1/2}
(1+\cos(2\pi \tilde \lambda )), \cr
&& \cr
T_{11} &=& - K_s K {\cal T}_{25}
\left[e^2(1-e^2)^{-1/2}(1+\cos(2\pi \tilde \lambda ))
+2 (1-e^2)^{1/2} f(\sqrt{1-e^2}x^0)\right], \cr
&& \cr
T_{ij} &=& -2K_s K {\cal T}_{25} ~(1-e^2)^{1/2}
f(\sqrt{1-e^2}x^0) \delta_{ij}, \cr
&& \cr
S_{01}&=& K_a K {\cal T}_{25} ~e(1-e^2)^{-1/2}
(1+\cos(2\pi \tilde \lambda ))\, .
\label{TS2}
\een

Before comparing these results with the ones 
obtained from the target space analysis in
section \ref{s2} (eqs.(\ref{e14})) we shall determine the constants $K_s$ 
and $K_a$
from the known results for the boundary state and the Dirac-Born-Infeld
action at $\tilde \lambda =0$.  The results for the sources obtained from
this boundary state can simply be derived by taking the $\tilde \lambda
\rightarrow 0$ limit of the results (\ref{TS2}). Now the corresponding
world-volume action takes the following standard form,
\be
\SSS_{DBI} = - {\cal T}_p \int d^{p+1}x \sqrt{-\det (g + b)}.
\label{SBI}
\ee
It is straightforward to compute the sources from this action. For
example, one gets the following results for $T_{00}$ and
$S_{01}$\footnote{As in sec. \ref{s2}, here also $S^{01}$ has
been defined in such a way that $\delta {\cal S}_{DBI}$ receives
contribution both from ${1\over 2}S^{01}\delta b_{01}$ and
${1\over 2}S^{10}\delta b_{10}$ .},
\ben
T_{00} = {\cal T}_p  ~(1-e^2)^{-1/2}, \qquad
S_{01} = -{\cal T}_p ~e (1-e^2)^{-1/2}.
\een
Comparing these with the corresponding results in (\ref{TS2}) in the
limit $ \tilde \lambda \rightarrow 0$ then fixes the constants to be,
\be \label{ekska}
K_s = {1\over 2K} \,, \qquad K_a = - {1\over 2K}\,.
\ee
With these values of the constants the results (\ref{TS2}) exactly match
with (\ref{e14}).

Given the ambiguity mentioned in the paragraph below \refb{e14}, one
might
wonder why the boundary state analysis leads precisely to the same 
expressions 
as \refb{e14}. In the
boundary state formalism 
the coupling of off-shell closed strings is defined by inserting the
corresponding
closed string vertex operator at the center of the disk.
To see why this prescription leads to \refb{e14}, note that if \refb{e12}
had held
for arbitrary time dependent $\delta g_{\mu\nu}$ and $\delta b_{\mu\nu}$,
then we could derive \refb{e14} explicitly using the target space 
analysis of section \ref{s2}. This would require that for
the relevant computations, the Seiberg-Witten equivalence relation
holds even when the background metric and anti-symmetric tensor fields
differ from constant values by an infinitesimal amount which depends on
the space-time coordinates. Since in the
boundary state analysis the $x^0$ dependence of various source terms (in
the Euclidean theory)
is computed using an insertion of a bulk operator 
$e^{ik.X^0(0)}$ at the center of
the disk, the agreement of boundary state analysis with \refb{e14} can be
explained if in
the computation of correlation functions of a single $e^{i k.X^0(0)}$ at
the
center,
and
open string vertex operators $\prod_i e^{ik_i.X^0(z_i)}$ inserted at the
boundary, we can continue to
use the  open string metric even though there is a bulk operator
insertion. This is indeed the case as long as there is only one insertion 
of the
bulk operator $e^{ik.X^0}$, since such a correlation function does not 
involve the bulk propagator, and the bulk-boundary propagator, like the 
boundary propagator, depends only on the open string 
metric\cite{9908142}.\footnote{In any case, as long as the operator 
$e^{ik.X^0}$ is inserted at the center of
the disk, the distance from the center of the disk to every point on 
the boundary is unity, and hence the correlator does not even involve the 
bulk-boundary propagator as it appears in the exponent of unity.}

\sectiono{Generalisation to superstrings}\label{superstrings}

The boundary CFT analysis  can be easily generalised to decay of non-BPS
D-branes
in
superstring theory\footnote{See \cite{nbps} and references therein.}
in the presence of an electric field $e$.  Here we shall
extract the sources for massless closed string fields from the boundary
state, as was done
in \cite{0203265}, and not attempt to
construct the full boundary state\footnote{See \cite{0108238} for
discussions on $SU(2)$ boundary states in the context of
superstrings.}.
Since the analysis is a straightforward extension of the results of the
previous section and \cite{0203265}, we shall only give the outlines
of
the derivation.

We work in the Wick rotated Euclidean theory as in the case of
bosonic string theory, and define the various coordinates $Y^a$,
$Z^a$ etc. in an identical manner. We also need to define the
fermionic partners of various coordinates, and in general we shall
denote the fermionic partners of $X^\mu$, $Y^a$ and $Z^a$ by
$\psi_x^\mu$, $\psi_y^a$ and $\psi_z^a$ respectively. The boundary
perturbation describing the Wick rotated rolling tachyon
background is given by, \be \label{ess1} \tl \int dt \, \,
\psi_z^0 \sin(Z^0/\sqrt 2) \otimes \sigma_1\, , \ee where
$\sigma_1$ is a Chan-Paton factor. Using manipulations similar to
those in bosonic string theory one can show that this can be
rewritten as \be \label{ess2} \tl \int dt \, \psi_{y L}^0 \,
\sin(\sqrt 2 Y^0_L)\otimes \sigma_1\, . \ee We can now construct
the boundary state by first taking $Y^0$ to be a compact
coordinate with radius $\sqrt 2$, and then recovering the result
for the non-compact case by throwing away all the winding modes.
For compact $Y^0$ coordinate we can represent the bosonic
coordinate $Y^0$ by a pair of fermionic coordinates $(\xi, \eta)$
satisfying the following relations:
 \ben \label{ess3} e^{i\sqrt 2
Y^0_L} &=& {1\over \sqrt 2} (\xi_L(t) + i \eta_L(t)) , \cr ~ \cr
e^{i\sqrt 2 Y^0_R} &=& {1\over \sqrt 2} (\xi_R(t) + i \eta_R(t))
\, . \een
 Then the operator (\ref{ess2}) reduces to
 \be \label{ereduces}
{1\over \sqrt{2}} \tl \int dt \psi^0_{yL} \eta_L\, ,
\ee
 and
produces a rotation
through angle $2\pi \tilde \lambda$ about the $\xi$ axis
\cite{0203265} on the unperturbed boundary state.

As in the case of bosonic string theory we proceed in three stages:
\begin{enumerate}
\item Construct the boundary state corresponding to the unperturbed
compactified BCFT ($\tl =0$)
including the background magnetic field $\bar b$.

\item Study the effect of switching on the perturbation $\tl$ by rotating
the boundary state by an angle $2\pi\tl$ about the $\xi$ axis.

\item Take the decompactification limit by removing all the winding modes
in the expression for the boundary state.

\end{enumerate}

For definiteness let us consider the non-BPS $\hbox{D} 9$-brane in
type IIA string theory. The corresponding boundary state in the
presence of magnetic field $\bbb$ but in the absence of any
perturbation $(\tl=0)$ is given by, \ben |\hbox{IIA}9 ; \bbb\ra
&=& |\hbox{IIA}9 ; \bbb, +\ra - |\hbox{IIA}9 ; \bbb, -\ra, \cr \cr
|\hbox{IIA}9 ; \bbb, \epsilon \ra &=& |\hbox{IIA}9 ; \bbb,
\epsilon \ra_{mat} \otimes  |\hbox{IIA}9 ; \bbb, \epsilon
\ra_{ghost}, \qquad \epsilon = \pm 1, \cr \cr |\hbox{IIA}9 ; \bbb,
\epsilon \ra_{mat} &=& N_\bbb \exp \left[-\sum_{n=1}^{\infty}
{1\over n} (\bet^0_{-n} \bar \gam^0_{-n} + \bet^1_{-n} \bar
\gam^1_{-n} + \sum_{j=2}^9 \alpha^j_{-n} \bar \alpha^j_{-n} )
\right. \cr \cr &&\left. - i\epsilon \sum_{r=1/2}^{\infty}
(\chi^0_{-r} \bar \delta^0_{-r} + \chi^1_{-r} \bar \delta^1_{-r} +
\sum_{j=2}^9 \psi^j_{-r} \bar \psi^j_{-r}) \right] |\hbox{IIA}9;
\bbb\ra_0 , \cr \cr |\hbox{IIA}9 ; \bbb\ra_0 &=& \sum_{m\in {\bf
Z}} \exp \left[ -i \sqrt 2 m (y^0_L -y^0_R) - i \sqrt 2 m \bar e
y^1 \right] |0\ra, \cr \cr |\hbox{IIA}9 ; \bbb, \epsilon
\ra_{ghost} &=& \exp \left[ - \sum_{n=1}^{\infty} \left(\bar
b_{-n}c_{-n} + b_{-n}\bar c_{-n} + i \epsilon (\bar \beta_{-n-1/2}
\gamma_{-n-1/2} - \beta_{-n-1/2} \bar \gamma_{-n -1/2})\right)
\right] |\Omega \ra, \cr \cr |\Omega \ra &=& (c_0 +c_0) c_1 \bar
c_1 e^{-\phi(0)} e^{-\bar \phi(0)} |0\ra, \cr \cr N_\bbb &=& iK
{{\cal T}_9 \over 2} (1+\bar e^2)^{1/2}\, , \label{IIA9} \een where
${\cal T}_9$ is the non-BPS $\hbox{D}9$-brane
tension\footnote{Note that the above state takes the same form for
the $\hbox{D}9-\bar{\hbox{D}}9$-brane system or the NS-NS part of
the BPS $\hbox{D}9$-brane in type IIB string theory. In each case
${\cal T}_9$ denotes the tension of the corresponding brane
system.\label{f1}}, $K$ is a convention dependent numerical factor
which does not affect the final result for the sources,
$\bet_n$'s are oscillators of $Y$, $\bar
\gam_{n}$'s are defined in (\ref{gamma}), $\{\chi_r, \bar
\chi_r\}$ and $\{\psi_r, \bar \psi_r\}$ are the sets of
oscillators of $\{\psi_y, \bar \psi_y \}$ and  $\{ \psi_x, \bar
\psi_x \}$ respectively, and
 \be
\bar \delta_r = M_Y \bar \chi_r. \label{delta} \ee
 {}From
eqs.(\ref{IIA9}) it is clear that the overall normalisation which
is the inner product between the NS-NS ground state corresponding
to the identity operator and the boundary state is $N_\bbb$. This
remains the overall normalisation for the rolling tachyon boundary
state as the rotation keeps the NS-NS vacuum invariant. Since the
effect of the magnetic field has been diagonalized to identity by
dealing with the oscillators $\bar \delta_r$, the state in
(\ref{IIA9})  looks, at least algebraically, exactly like a
Neumann boundary state which was considered in \cite{0203265}. The
only difference is an extra factor of $y^1$-momentum dependence
for each $y^0$ winding state.

As discussed already, the effect of the perturbation \refb{ess1}
is an SO(3) rotation about the $\xi$-axis by an angle $2\pi\tl$ on
the boundary state. The effect of this rotation on the unperturbed
boundary state can be easily studied following \cite{0203265}.
Using the result $Y^0-\bar e Y^1 = \sqrt{1+\bar e^2} X^0$, the
result for the part of the perturbed boundary state which involves
either no oscillators or the states created by the action of
$\chi^0_{-1/2}\bar \delta^0_{-1/2}$ on pure momentum states is
given by\footnote{Notice that the coefficients of
$\chi^0_{-1/2}\bar \delta^0_{-1/2}$ terms in eqs. (\ref{part1})
and (\ref{part2}) have been sign flipped with respect to the
corresponding terms in
 eqs. (4.6) and (4.7) respectively in \cite{0203265}. This is because
these equations in \cite{0203265} were written after performing
the inverse Wick-rotation, while here we are still in the
Wick-rotated theory.} ,
 \be
N_\bbb \left[ \wh f\left(\sqrt{1+\bar
e^2} X^0(0)\right) - i \eps \chi^0_{-1/2}\bar \delta^0_{-1/2}~ \wh
g\left( \sqrt{1+\bar e^2} X^0(0) \right) \right] |0\ra,
\label{part1} \ee
 where
 \ben \label{defbarf} \wh f(x) &=& \left[ 1
~+~ (-1)^n \sum_{n=1}^{\infty} (\sin(\tilde \lambda \pi) )^{2n}
\left( e^{in\sqrt{2} x} +
e^{-in\sqrt{2} x} \right) \right] = f(-ix) \, , \nonumber \\
\wh g(x) &=& 1 + \cos(2\pi\tl) - \wh f(x)\, ,
\een
 with $f(x)$
given in (\ref{exx1}) for the superstring case. Therefore the
net level (1/2, 1/2) contribution to the rolling tachyon boundary
state is \cite{0203265}, \ben &&K{\cal T}_9 (1+\bar
e^2)^{1/2} \Bigg[ \chi^0_{-1/2}\bar \delta^0_{-1/2} ~ \wh
g\left(\sqrt{1+\bar e^2} X^0(0)\right) \cr &&\cr &&+ \left(
\chi^1_{-1/2}\bar \delta^1_{-1/2} + \sum_{j=2}^9 \psi^j_{-1/2}
\bar \psi^j_{-1/2} \right) \wh f\left(\sqrt{1+\bar e^2}
X^0(0)\right) \cr &&\cr &&+( \bar \beta_{-1/2} \gamma_{-1/2} -
\beta_{-1/2} \bar \gamma_{-1/2} ) \wh f\left(\sqrt{1+\bar e^2}
X^0(0)\right) \Bigg] |\Omega \ra. \label{part2} \een 
Now if we write the above state
after inverse Wick-rotation and analytic
continuation (\ref{analytic}) in the following form,
 \be -\int
d^{10}k \left[ (\tilde A_{\mu \nu}(k) + \tilde C_{\mu \nu}(k))
\psi^{\mu}_{-1/2} \bar \psi^{\nu}_{-1/2} + \tilde B(k) (\bar
\beta_{-1/2} \gamma_{-1/2} - \beta_{-1/2}\bar \gamma_{-1/2} )
\right] |\Omega, k\ra, \label{part3} \ee
 with $\tilde A_{\mu \nu}$
symmetric and $ \tilde C_{\mu \nu}$ anti-symmetric, then the
Fourier transforms are given by the same equations as in (\ref{B})
and (\ref{ASprime}) with ${\cal T}_{25}$ replaced by ${\cal T}_9$
and $f(x)$ given by \refb{exx1}. Identifying the level $(1/2,
1/2)$ part of the relation $(Q_B + \bar Q_B) |\BB\ra=0$ with the
conservation law $\p^\mu T_{\mu\nu}=\p^\mu S_{\mu\nu}=0$, we arrive at
the same result as \refb{TS}.
This, in turn, leads to the same equations as \refb{TS2} with $f(x)$
given in \refb{exx1} and $\TT_{25}$ replaced by $\TT_{9}$. Note
that
${\cal T}_9$ in this case has to be interpreted as the net tension
of whatever D-brane system we consider (see footnote \ref{f1}). To
compute the constants $K_s$ and $K_a$ one can proceed in a similar
way through the Dirac-Born-Infeld action as was done in the previous
section and one arrives at the same results as in \refb{ekska}.
This gives the expected results for the sources as given in
eq.(\ref{e14}).

\sectiono{Effective Field Theory} \label{s4}

Although in section \ref{s2} we used the existence of an effective
field theory in describing the dynamics of tachyon condensation,
we did not commit ourselves to any particular choice of the
effective action. In this section we shall discuss a specific form
of the low energy effective action that reproduces the answers
obtained in sections \ref{s2}-\ref{superstrings} at late time.
This action is conjectured to describe correctly the classical open string
dynamics
when the second and higher derivatives of the tachyon and all other gauge
and massless scalar fields are small.
For definiteness we
shall focus our attention on non-BPS D-branes, but the results can
be easily generalised to brane-antibrane system as well.

The proposed action (following \cite{9909062,0009061,effective,0003221})
for describing the
dynamics of the
tachyon $T$ and gauge
fields $A_\mu$ on a D-$p$-brane in the presence of constant background
metric
$g_{\mu\nu}$, anti-symmetric tensor field $b_{\mu\nu}$ and dilaton $\phi$
is:
\be \label{ex2}
S = -\int d^{p+1} x\, e^{-\phi} \,  V(T) \, \sqrt{-\det A}\, ,
\ee
where\footnote{Note that this choice of the potential $V(T)$ is in
apparent
contradiction with the potential derived in boundary string field
theory\cite{bsfttach}. This paradox disappears if we note that $T$
appearing in \refb{ex2}, \refb{ex3} could be related to the tachyon field
of boundary string field theory via a complicated field redefinition which
includes derivative terms. For example, an action
of
the form $-\int d^{p+1}x (\eta^{\mu\nu} \p_\mu \phi \p_\nu \phi + V(\phi)
+ \ldots)$ where $\ldots$ denote terms involving higher powers of
derivatives, can be transformed to an action of the form $-\int d^{p+1}x
(\eta^{\mu\nu} \p_\mu \psi \p_\nu \psi + U(\psi)
+ \ldots)$ with a different form of the potential $U(\psi)$, by a field
redefinition of the form $\phi = f(\psi) + g(\psi) \p^\mu\psi \p_\mu\psi +
\ldots$ by appropriately choosing the functions $f$ and $g$.}
\be \label{ex3}
V(T) \simeq e^{-\alpha T / 2}\, ,\qquad \hbox{for large $T$}\, ,
\ee
\be \label{ex3a}
A_{\mu\nu} = g_{\mu\nu} + b_{\mu\nu} + F_{\mu\nu}
+ \partial_{\mu} T \partial_{\nu} T + \p_\mu Y^m \p_\nu Y^m\, ,
\ee
\be \label{ex4}
F_{\mu\nu} = \p_\mu A_\nu - \p_\nu A_\mu\, ,
\ee
and $Y^m$ ($m\ge p+1$) denote coordinates transverse to the D-$p$-brane.
$\alpha=1$ for bosonic string theory and $\sqrt 2$ for superstring theory.

We shall now show that this action reproduces the correct time dependence
of
the energy-momentum tensor $T_{\mu\nu}$, and the source $S_{\mu\nu}$ for
the anti-symmetric tensor field associated with rolling tachyon solution
at
late time. For this we note that $T_{\mu\nu}$ and $S_{\mu\nu}$
computed from this action are given by:
\ben \label{ex5}
T_{\mu\nu} &=& -{1\over 2} e^{-\phi} \, V(T) \sqrt{-\det A} \, \{
(A^{-1})_{\mu\nu}
+ (A^{-1})_{\nu\mu}\}\, , \nonumber \\
S_{\mu\nu} &=& -{1\over 2}e^{-\phi} \, V(T) \sqrt{-\det A} \, \{
(A^{-1})_{\mu\nu}
- (A^{-1})_{\nu\mu}\}\, .
\een
For $\phi=0$,
$A_\mu=0$, $Y^m=0$,
background $g_{\mu\nu}$, $b_{\mu\nu}$ of the form \refb{e2},
and spatially homogeneous tachyon field, $T_{\mu\nu}+S_{\mu\nu}$ is given
by the
following matrix:
\be \label{ex6}
{V(T) \over \sqrt{1 - e^2 - (\p_0 T)^2}} \, \pmatrix{ 1 & -e & \cr e &
-e^2
& \cr  &  & \zero_{p-1}}
- V(T) \,  \sqrt{1 - e^2 - (\p_0 T)^2}\, \pmatrix{ 0 & 0 & \cr 0 & 1 & \cr
& & \one_{p-1}}\, .
\ee
Since
$T_{00}= V(T)/\sqrt{1 - e^2 - (\p_0 T)^2}$ must
be conserved, we see that as $T$ rolls towards $\infty$, $\p_0 T$ must
approach $\sqrt{1-e^2}$. Thus in this limit:
\be \label{ex7}
T \simeq \sqrt{1-e^2}\, x^0 + C, \qquad
V(T) \simeq e^{-\alpha C/2 -  \alpha\sqrt{1-e^2}\, x^0/2}, \qquad
\sqrt{1-e^2 - (\p_0 T)^2} \simeq K e^{-  \alpha\sqrt{1-e^2}\, x^0/2}\, ,
\ee
so that,
\ben \label{ex8}
T_{00} &=& K^{-1} \, e^{-\alpha C/2}\, , \nonumber \\
T_{01} &=& 0\, , \nonumber \\
T_{11} &\simeq& -e^2 K^{-1} \, e^{-\alpha C/2} - K \, e^{-\alpha C/2} e^{-
\alpha\sqrt{1-e^2}\, x^0}\, , \nonumber \\
T_{ij} &\simeq& -K \, e^{-\alpha C/2} e^{-
\alpha\sqrt{1-e^2}\, x^0}\,\delta_{ij}\, , \qquad \hbox{for} \quad i,j\ge
2\, , \nonumber \\
S_{01} &=& -e K^{-1} \, e^{-\alpha C/2}\, , \een
with all other
components of $T_{\mu\nu}$ and $S_{\mu\nu}$ being zero. This
reproduces the large time behaviour of \refb{e14} provided we
identify $K^{-1} \, e^{-\alpha C/2}$ with $(1-e^2)^{-1/2} \TT_p (1
+ \cos(2\pi\tl))/2$ and $K  e^{-\alpha C/2}$ with the coefficient
of $e^{-\alpha\sqrt{1-e^2} x^0}$ in the large $x^0$ behaviour of
$(1-e^2)^{1/2}\TT_p f(\sqrt{1-e^2}x^0)$.

As discussed in \cite{0009061,0010240,0204143}, the dynamics of the system
described by
the
effective action \refb{ex2} is best described in the Hamiltonian
formulation. Since this has been studied in detail in these papers, we
shall not discuss it here.

In the presence of the background RR $p$-form field $C^{(p)}$ there is an
additional coupling of the form\cite{nbpscs}:
\be \label{eadd1}
\int d^{p+1} x f(T) \, dT \wedge C^{(p)}\, ,
\ee
where
\be \label{eadd2}
f(T) \simeq  e^{- T/\sqrt 2}\, ,
\ee
for large $T$\cite{0204143}.
The source of the RR $p$-form field, computed from this term, has the
correct time dependence\cite{0204143}.
If there are background $q$-form RR
fields $C^{(q)}$ for $q<p$ as well, then \refb{eadd1} is generalised to
\cite{csterm}
\be \label{eadd3}
\int d^{p+1} x f(T) \, dT \wedge e^{b+F} \wedge C\, ,
\ee
where
\be \label{eadd4}
C = \sum_{q\le p} C^{(q)}\, .
\ee

For completeness of
our discussion we shall briefly review the coupling of the
field theory action \refb{ex2} to background
supergravity fields following
\cite{oldsup,9611173,9909062,0003221}.
We define:
\be \label{esup1}
\GG_{\mu\nu} = \p_\mu Z^M \p_\nu Z^N E_M^a E_N^b \eta_{ab}\, ,
\ee
where $Z^M$ ($M$ running over 10 bosonic and 32 fermionic coordinates) are
the
D-brane world-volume fields describing superspace
coordinates, $\p_\mu$ denotes derivative with respect to the D-brane
world-volume coordinates $x^\mu$ and $E_M^A$ (with $A$ running over 10
bosonic
and 32 fermionic
coordinates) are the bulk fields denoting the supervielbeins.  The
index $a$
runs over the subset of 10
bosonic
indices.
We can choose the static gauge where the bosonic components
$Z^\mu$ of $Z^M$ are set equal to the D-brane world-volume coordinates
$x^\mu$, but we shall not choose any specific gauge here. We also define:
\be \label{esup2}
\FF_{\mu\nu} = \p_\mu A_\nu - \p_\nu A_\mu + B_{MN} \p_\mu Z^M \p_\nu Z^N
\, ,
\ee
\be \label{exy1}
\AAA_{\mu\nu} = \GG_{\mu\nu} + \FF_{\mu\nu} + \p_\mu T \p_\nu T\, ,
\ee
and
\be \label{esup2a}
\CC^{(q)}_{\mu_1\ldots \mu_q} = C^{(q)}_{M_1\ldots M_q} \p_{\mu_1} Z^{M_1}
\ldots \p_{\mu_q} Z^{M_q}\, ,
\ee
where  $B_{MN}$ is the NS-NS two form field of supergravity in superspace,
$C^{(q)}_{M_1\ldots M_q}$ is the RR $q$-form field of supergravity in
superspace,
and $A_\mu$ is the U(1) gauge field on the D-brane world-volume.
Finally
let $\phi$ denote the dilaton field in the bulk.
Then according to the results of
\cite{9909062,0003221,9908142}, the coupling of the action
\refb{ex2} to the supergravity background will be given by:
\be \label{esup4}
S = - \int d^{p+1} x \, V(T) \, e^{-\phi} \, \sqrt{-\det(\AAA)}
+ \int d^{p+1} x \, f(T) \, d T \wedge \, \sum_{q\le p} \CC^{(q)} \wedge
e^\FF \, .
\ee
This completes our discussion of the coupling of the supergravity fields
to
the tachyon effective field theory.

\bigskip

{\bf Acknowledgement}: We would like to thank S.~Das, D.~Ghoshal, 
R.~Gopakumar, 
D.~Jatkar, J.~Maldacena, J. Michelson, S.~Minwalla, S.~Naik, T.~Okuda,
S.~Panda,
O.~Pavlyk, L.~Rastelli, A.~Strominger, S.~Sugimoto and B.~Zwiebach for
useful
discussions. We thank A. Recknagel
for useful communication. P.M. would like to acknowledge the support of
Penn State University through a Teaching Assistantship during part of this
work. He also acknowledges the hospitality of the Physics Department of
Kentucky University where a part of this work was done. The research of
A.S. was supported in part by a grant from the Eberly College of Science
of the Penn State University. A.S. would also like to acknowledge the
hospitality of the YITP at Stony Brook, Center for Theoretical Physics at
MIT, and a grant from the NM Rothschild and Sons Ltd at the Isaac Newton
Institute where part of this work was done.

\appendix

\sectiono{Divergence of the Boundary State in the $e\to 1$, $\tl\to 1/2$
Limit} \label{divergence}

In this section we shall discuss the behaviour of the boundary
state in the limit (\ref{elimit}). From equations (\ref{e14}) it
is clear that the sources of all the massless closed string states
are finite. But we shall see here that the sources for higher
massive states diverge in this limit. We shall demonstrate this by
computing the next higher level terms in the boundary state. In
particular we shall focus on the state, \ben |\BB_{\bbb,\tl}
\ra_{Y^0,Y^1} &=& N_\bbb \sum_{j, m} D^j_{m, -m} |j, m, m \ra
\ra^{(0)}_{\bar \gam} \otimes \exp\left[ -\sum_{n>0} {1 \over n}
\bet^1_{-n} \bar \gam^1_{-n} -2im \bar e y^1 \right] |0\ra,
\label{BY} \een which is the part of $|\BB_{\bbb,\tl} \ra$ given
in (\ref{Blambda-bs3}) involving $Y^0$ and $Y^1$ . The level $(2,2)$
oscillator part of this state is given by,
 \be |\BB_{\bbb,
\tl}\ra_{Y^0,Y^1}^{(2)}  = N_\bbb \left[ |\BB_2\ra - \bet^1_{-1}
\bar \gam^1_{-1} |\BB_1\ra - {1\over 2} \left(\bet^1_{-2} \bar
\gam^1_{-2} - (\bet^1_{-1} \bar \gam^1_{-1})^2  \right)|\BB_0\ra
\right]\, , \label{BY2} \ee
 where the states $|\BB_2\ra $, $|\BB_1\ra
$ and $ |\BB_0\ra $ are the oscillator level (2,2), (1,1) and
(0,0) parts respectively of the state, \ben |\BB \ra &=& \sum_{j,
m} |D;j,m \ra \otimes \exp\left[ -2im \bar e y^1 \right] |0\ra,
\cr &&\cr |D;j,m \ra &\equiv& D^j_{m, -m} |j, m, m \ra
\ra^{(0)}_{\bar \gam} \, .\label{calB} \een The oscillator
contribution comes only from $|D;j,m \ra $. Let us first consider
the state $|\BB_2 \ra$. To isolate the relevant parts, we note
that $|j,m,m\ra$ has conformal weight $(j^2,j^2)$, of which
$(m^2,m^2)$ comes from the $Y^0$ momentum and $(j^2-m^2, j^2-m^2)$ comes
from the oscillators. {}From this we see that
the contribution to $|\BB_2 \ra$ comes from three different types
of terms.
\begin{enumerate}
\item Level (2,2) secondary in $|D;j,\pm j\ra$. This contribution
can be further divided into three different parts. \ben
\label{first} j\ge 1 &:& \quad \frac{1}{2}(-\sin(\tl\pi))^{2|j|}
\{ (\bet^0_{-1})^2 (\bar\gam^0_{-1})^2 +
\bet^0_{-2}\bar\gam^0_{-2}\} \exp(\pm 2ij Y^0(0))|0\ra\, , \nonumber \\
j={1\over 2} &:& \quad (-\sin(\tl\pi)) \, {1\over 6} \,
\{(\bet^0_{-1})^2 \pm \sqrt 2 \, \bet^0_{-2}\} \, \{
(\bar\gam^0_{-1})^2 \pm \sqrt 2 \, \bar\gam^0_{-2}\} \exp(\pm
iY^0(0))|0\ra \, , \nonumber \\
j=0 &:& \quad {1\over 2}\,(\bet^0_{-1})^2 (\bar\gam^0_{-1})^2
|0\ra\, .  \een

\item Level (1,1) secondary in $|D;1,0\ra$. This is given by:
\be \label{second} -{1\over 2} \cos(2\tl\pi)
\,\bet^0_{-2}\bar\gam^0_{-2}) |0\ra\, . \ee

\item The primary state in $|D;{3\over 2}, \pm{1\over 2}\ra$. This
contribution is given by: \be \label{third} \sin(\tl\pi)
(\cos(2\tl\pi)+\cos^2(\tl\pi)) \, {1\over 6} \, \{\sqrt
2(\bet^0_{-1})^2 \mp \, \bet^0_{-2}\} \, \{ \sqrt
2(\bar\gam^0_{-1})^2 \mp \, \bar\gam^0_{-2}\} \exp(\pm
iY^0(0))|0\ra \, .\ee
\end{enumerate}
Note that whereas the phases of the states $|D;j,\pm j\ra$ and
$|D;1,0\ra$ were determined in \cite{0203211,0203265}, the phase
of $|D;{3\over 2}, \pm{1\over 2}\ra$ needs to be determined afresh
by requiring that at $\tl={1\over 2}$ the boundary state reduces
to the known boundary state. This will be checked explicitly
later.

Combining these results together, we get the following final
expression for  $|\BB_2\ra$,
 \ben |\BB_2\ra &=& \Bigg[ \bet^0_{-2}
\bar \gam^0_{-2} B^{(2)}\left( \tl; X^0(0) \right) +
(\bet^0_{-1})^2 (\bar \gam^0_{-1})^2 B^{(4)}\left( \tl; X^0(0)
\right) \cr && + \left( \bet^0_{-2} (\bar \gam^0_{-1})^2 +
(\bet^0_{-1})^2 \bar \gam^0_{-2} \right) B^{(3)}\left( \tl; X^0(0)
\right) \Bigg] |0\ra \label{BB2}, \een where, \ben B^{(2)}\left(
\tl; X^0(0) \right) &=& -{1\over 2} \cos(2\pi \tl)
 +  \sin(\pi \tl) \cos^2(\pi \tl) \cos
 \left(\sqrt{1+\bar e^2}
 X^0(0)\right) \cr
&& \cr && + {1\over 2} \left(\wh f\left( \sqrt{1+\bar e^2}
X^0(0)\right) -1\right), \cr && \cr B^{(3)}\left( \tl; X^0(0)
\right) &=&  -{i }\sqrt{2} \sin(\pi \tl) \cos^2(\pi \tl)
\sin\left( \sqrt{1+\bar e^2} X(^00)\right), \cr &&\cr
B^{(4)}\left( \tl; X^0(0) \right) &=& 2 \sin(\pi \tl) \cos^2(\pi
\tl) \cos\left( \sqrt{1+\bar e^2} X^0(0)\right) + {1\over 2} \wh
f\left(\sqrt{1+\bar e^2} X^0(0)\right), \cr && \label{Bcoeff} \een
where the functions $\wh f(x)$, $\wh g(x)$ were defined in 
eq.(\ref{edeffbgb}).

The results for $|\BB_1\ra$ and $|\BB_0\ra $ are already known
from the analysis of \cite{0203265} which was also discussed in
sec. \ref{sources}. These are given by, \be |\BB_1\ra = -
\bet^0_{-1} \bar \gam^0_{-1} ~\wh g\left( \sqrt{1 +\bar e^2}
X^0(0)\right) |0\ra, \qquad |\BB_0 \ra = \wh f\left(\sqrt{1+\bar
e^2} X^0(0)\right) |0\ra \, . \label{BB10} \ee 
Collecting all these results for $|\BB_2\ra,
|\BB_1\ra $ and $|\BB_0\ra$ and substituting them in
eq.(\ref{BY2}) one finally gets,
 \ben
|\BB_{\bbb,\tl}\ra^{(2)}_{Y^0,Y^1} &=& N_\bbb \left[
\bet^0_{-2}\bar \gam^0_{-2} \left\{  \sin(\pi \tl) \cos^2(\pi \tl)
\cos\left(\sqrt{1+\bar e^2} X^0(0) \right) - {1\over 2} \wh
g\left(\sqrt{1+\bar e^2} X^0(0)\right) \right \} \right. \cr &&\cr
&+& (\bet^0_{-1} \bar \gam^0_{-1})^2 \left\{  2 \sin(\pi \tl)
\cos^2(\pi \tl) \cos\left(\sqrt{1+\bar e^2} X^0(0) \right) +
{1\over 2} \wh f\left(\sqrt{1+\bar e^2} X^0(0)\right) \right\} \cr
&&\cr
 & -& {i }\sqrt{2} \left( \bet^0_{-2} (\bar \gam^0_{-1})^2 +
(\bet^0_{-1})^2
 \bar \gam^0_{-2} \right)
\sin(\pi \tl) \cos^2(\pi \tl) \sin\left(\sqrt{1+\bar e^2} X^0(0)
\right) \cr &&\cr &+& \bet^0_{-1}\bet^1_{-1}\bar
\gam^0_{-1} \bar \gam^1_{-1} ~\wh g\left(\sqrt{1+\bar e^2}
X^0(0)\right) \cr &&\cr
&-&  \left. {1\over 2} \left( \bet^1_{-2}\bar
\gam^1_{-2} -
(\bet^1_{-1} \bar \gam^1_{-1} )^2 \right) \wh f\left(\sqrt{1+\bar
e^2} X^0(0)\right) \right ]|0\ra, \label{eabc}\een

One can now use the matrices $v$ and $M_Y$ given in equations
(\ref{Vv}) and (\ref{eas7}) to express \refb{eabc} in terms of the
$\alpha$ and $\bar \alpha$ oscillators. To exhibit the divergence
of the boundary state we shall state the result for the
coefficient of the state $(\alpha^0_{-1} \bar \alpha^0_{-1})^2
|0\ra$. This is given by, \be {K\, {\cal T}_p }(1+\bar
e^2)^{-3/2} \left[(1+ \cos(2\pi \tl)) \left(\sin(\pi \tl)
\cos(\sqrt{1+\bar e^2} x^0) -  \bar e^2 \right) +
 \frac{1}{2}  (1+\bar e^2)^2 \wh f(\sqrt{1+\bar e^2}
x^0)\right]. \label{bcd} \ee
After inverse Wick-rotation and the analytic continuation
(\ref{analytic}) this becomes,
 \be {K\, {\cal T}_p} (1-
e^2)^{-3/2} \left[(1+ \cos(2\pi \tl)) \ \left( \sin(\pi \tl)
\cosh(\sqrt{1- e^2} x^0) +  e^2 \right) + \frac{1}{2} (1-e^2)^2
f(\sqrt{1-e^2} x^0)\right], \label{coeff} \ee
 In the limit
(\ref{elimit}), the first term in the above expression diverges
while the second term goes to zero. The first term diverges as
$\lim_{e\rightarrow 1} 4K \rho (1-e^2)^{-1}$, where $\rho =
\displaystyle{ {{\cal T}_p \over 2}} (1- e^2)^{-1/2}(1+\cos(2\pi
\tl))$ is the energy density $T^{00}$ in this limit and it is
finite.  We can also verify that at $\tl = 1/2$, $e<1$, the  state
\refb{coeff} vanishes identically, as is required by the fact that
at finite $e$, $\tl=1/2$ represents the tachyon vacuum
configuration.

\end{document}